\documentclass[12pt,letterpaper]{article}
\usepackage[latin1]{inputenc}
\usepackage{amsmath}
\usepackage{amsfonts}
\usepackage{amssymb}
\usepackage{graphicx}
\usepackage[left=1.25in,right=1.25in,top=1in,bottom=1in]{geometry}

\usepackage{setspace}
\newcommand{\no}[1]{}

\usepackage{amsmath,amsfonts,amssymb}
\usepackage{graphicx}
\usepackage{ifthen}
\usepackage{listings}
\usepackage{courier}
\usepackage{pstricks,pst-node,pst-tree}
\usepackage{float}
\usepackage{algorithm}
\usepackage{algorithmic}

\newtheorem{theorem}{Theorem}

\long\def\symbolfootnote[#1]#2{\begingroup%
\def\thefootnote{\fnsymbol{footnote}}\footnote[#1]{#2}\endgroup}

\begin{document}

\begin{center}
\Large{\bf{Faster Compact Top-k Document Retrieval\symbolfootnote[1]{Funded by Fondecyt grant 1-110066 and by the Conicyt PhD Scholarship
Program, Chile.}}}\\
\vspace{0.5cm}
\normalsize{Roberto Konow and Gonzalo Navarro}\\
\vspace{0.25cm}
\em{Department of Computer Science, University of Chile \\}
\small{{\{rkonow,gnavarro\}@dcc.uchile.cl}}

\end{center}

\normalsize

\begin{abstract}
\noindent \textit{\textbf{Abstract:}} An optimal index solving top-$k$ document
retrieval [Navarro and Nekrich, {\em SODA'12}] takes $O(m+k)$ time for a 
pattern of length $m$, but its space is at least $80n$ bytes for a collection 
of $n$ symbols. We reduce it to $1.5n$--$3n$
bytes, with $O(m + (k+\log \log n)\log \log n)$ time, on typical texts. 
The index is up to 25 times faster than the best previous compressed solutions,
and requires at most 5\% more space in practice (and in some cases as
little as one half). Apart from replacing
classical by compressed data structures, our main idea is to
replace {\em suffix tree sampling} by {\em frequency thresholding} to achieve 
compression.
\end{abstract}

\section{Introduction}

Finding the $k$ most documents relevant to a query is at the heart of search 
engines and information retrieval \cite{BU10}. A simple relevance 
measure is the number of occurrences of the query in the documents ({\em term 
frequency}). Typically the data structure employed to solve those ``top-$k$'' 
queries is the \emph{inverted index}. Inverted indexes work well, but they are 
limited to scenarios where the queryable terms are predefined and not too many
(typically ``words'' in Western languages), while they cannot search for 
arbitrary patterns (i.e., substrings in the sequences of symbols). This 
complicates the use of inverted indexes for Oriental languages such as Chinese,
Japanese and Korean, for agglutinating languages such as Finnish and German, 
and in other types of collections containing DNA and protein sequences, source 
code, MIDI streams, and other symbolic sequences. Top-$k$ document retrieval 
is of interest on those more general sequence collections 
\cite{M02,HSV09,NN12,NV12}, yet the problem of finding top-$k$ documents 
containing the pattern as a substring, even with a simple measure like term 
frequency, is much more challenging.

The general problem can be defined as follows: Preprocess a collection of $d$ 
documents containing sequences of total length $n$ over an alphabet of size 
$\sigma$, so that later, given a query string $P$ of length $m$, one retrieves 
$k$ documents with highest relevance to $P$, for some definition of relevance. 
Hon et al.~\cite{HSW09,HPSW10} presented the first efficient solution for this 
problem, achieving $O(m+\log n \log \log n+k)$ time, yet with super-linear 
space usage, $O(n\log^2n)$ bits. Then Hon et al.~\cite{HSV09} improved the 
solution to $O(m+k\log k)$ time and linear space, $O(n\log n)$ bits. Recently, 
Navarro and Nekrich \cite{NN12} achieved optimal $O(m+k)$
time, using $O(n(\log \sigma+\log d))$ bits.
Although the latter solution essentially closes the problem in theoretical terms, the constants involved are not small, especially in space: Their index can use up to $80n$ bytes, making it unfeasible for real scenarios.

There has been some work aiming to reduce the space of top-$k$ indexes
\cite{CNPTesa10,BN11,NV12}, yet they come at the cost of search times of at 
least $O(m+k\log k \log^{1+\epsilon} n)$ for any constant $\epsilon>0$,
while reaching as low as $n\log\sigma + O(n\log\log\log d)$ bits of space
(all our logarithms are in base 2).
In practice, the best ones \cite{NV12} require $2n$ to $4n$ bytes and answer
top-10 queries in about a millisecond. 
Their main idea is {\em suffix tree sampling}, that is, store the top-$k$
answers for large enough suffix tree nodes.

Hon et al.~\cite{HST12} have proposed an intermediate alternative, which is
basically an engineered implementation of their classical scheme \cite{HSV09}.
They use $(\log\sigma+2\log d)(n+o(n))$ bits
and $O(m+k\log\log n+(\log\log n)^4)$ time, or $(\log\sigma+\log d)(n+o(n))$ bits
and $O(m+k(\log\sigma\log\log n)^{1+\epsilon} + (\log\log n)^6)$ time.
This solution has not been yet implemented, however. We estimate their space 
would be at least $4n$ to $6n$ in practice.

In this work we design and implement a fast and compact solution for top-$k$
document retrieval, building on the ideas of Navarro and Nekrich \cite{NN12}. 
Apart from replacing classical by compact data structures, we use a novel idea
of {\em frequency thresholding} instead of sampling suffix tree nodes: We store
all the solutions for all the suffix tree nodes, but discard those with 
frequency 1.

We obtain time $O(m +(k+\log \log n)\log \log n)$ and space
$(\log\sigma+\log d+4\log\log n)(n+o(n))$ bits for typical texts.
By ``typical'' we mean that our results
hold almost surely (a.s.\footnote{A sequence $X_n$ tends to a value $\beta$ 
almost surely if, for every $\epsilon>0$, the probability that 
$|X_N/\beta-1|>\epsilon$ for some $N>n$ tends to zero as $n$ tends to infinity, 
$\lim_{n\rightarrow \infty}\sup_{N>n} \mathrm{Pr}(|X_N/\beta-1|>\epsilon)=0$.},
a very strong kind of convergence) for texts sampled from a stationary mixing
ergodic source (more precisely, type A2 in Szpankowski's sense \cite{Szp93}).
This is also a quite general assumption including Bernoulli and Markovian
models.

In addition, we have implemented our index, showing its practicality. It
turns out to require about $1.5n$--$3n$ bytes, that is, 25--50 times less than a naive
implementation of the basic idea \cite{NN12} and at most 5\% more space than
the most compressed practical solutions \cite{NV12} (while in some cases our index
uses half the space). Its time per query is $k$--$4k$
microseconds, outperforming the more compressed solutions by up to 25 times.
This is the first top-$k$ index for general texts 
that achieves little space and microseconds-time.
Moreover, it shows that our idea of thresholding frequencies generally gives
better results than the previous trend of sampling suffix tree nodes.

\section{Basic Concepts}

Consider a collection of string documents $D_i$ as the concatenation $T[1,n] =
D_1,D_2 \dots$ $D_d$, $T = t_1 t_2 \ldots t_n$, where at the end of each $D_i$ a 
special symbol \$ is used to mark the end of that document.
A {\em suffix array} \cite{MM93} $SA[1,n]$ contains pointers to every suffix of
$T$, lexicographically sorted. For a position $i \in [1,n]$, $SA[i]$
points to the suffix $T[SA[i],n] = t_{SA[i]}t_{SA[i]+1}\dots t_n$, where it
holds $T[SA[i],n] < T[SA[i+1],n]$. The $occ=ep-sp+1$ occurrences of a pattern $P[1,m]$
in $T$ are pointed from a range $SA[sp,ep]$, that can be found and listed in 
time $O(m\log n+occ)$.

The {\em suffix tree} \cite{W73} of $T$ is a path-compressed trie (i.e., unary paths are
collapsed) in which all the
suffixes of $T$ are inserted. Internal nodes correspond to repeated strings of
$T$ and the leaves correspond to suffixes. For internal nodes $v$, $path(v)$
is the concatenation of the edge labels from the root to $v$. The suffix tree 
finds the occurrences of $P$ in $T$ in time
$O(m+occ)$, by traversing it from the root to the {\em locus} of $P$, i.e.,
the highest node $v$ such that $P$ is a prefix of $path(v)$. Then all the 
occurrences of $P$ correspond to the leaves of the subtree rooted at $v$.
These leaves correspond to the range $SA[sp,ep]$, indeed, 
$v$ is the lowest common ancestor of the $sp$th and the $ep$th leaves. 
The suffix tree has $O(n)$ nodes.

{\em Compressed suffix
arrays (CSAs)} \cite{NM07} can represent the text {\em and} its suffix array
within essentially $nH_k(T) \le n\log\sigma$ bits. Here $H_k(T)$ is the 
empirical $k$th order entropy of $T$ \cite{Man01}, a lower bound to the bits 
per symbol emitted by a statistical compressor of order $k$. This
representation allows us to {\em count} (determine the interval $[sp,ep]$
corresponding to a pattern $P$), {\em access} (compute $SA[i]$ for any $i$),
and {\em extract} (rebuild any $T[l,r]$). We use one \cite{BNesa11} that can 
count in time $O(m)$, access in time $O(s)$ and extract in
time $O(s+r-l)$ while using $nH_k(T)+o(nH_k(T))+O(n+(n/s)\log n)$ bits for
any $k \le \alpha\log_\sigma n$ and any constant $0<\alpha<1$. The 
$O((n/s)\log n)$ bits correspond to storing one $SA[i]$ value every $s$
text positions. 

{\em General trees} of $n$ nodes can be represented using $2n+o(n)$ bits.
In this paper we use a representation \cite{SN10} that supports in 
$O(1)$ time a number of operations, including $preorder(v)$ (the 
preorder of node $v$), $preorderselect(i)$ (the $i$th node in preorder),
$depth(v)$ (depth of node $v$), $subtreesize(v)$ (number of nodes in subtree
rooted at $v$), $lca(u,v)$ (lowest common ancestor of nodes $u$ and $v$), and
many others. This structure is practical and implemented \cite{ACNS10}, 
using $2.37n$ bits. 

{\em Bitmaps} $B[1,n]$ can be represented using $n+o(n)$ bits, so that 
we can solve in constant time operations $rank_b(B,i)$ (number of occurrences
of bit $b$ in $B[1,i]$) and $select_b(B,j)$ (position in $B$ of the $j$th
occurrence of bit $b$) \cite{Mun96}. We use an implementation \cite{GGMN05}
that requires $1.05n$ bits, yet for very sparse bitmaps (with $m<<n$ bits set)
we prefer a compressed one using $m\log(n/m)+2m$ bits \cite{OS07}.

{\em Range Maximum Queries (RMQs)} ask for the position of the maximum element
in a range of an array, $\textsc{rmq}_A(i,j) = \mathrm{argmax}_{i\le k\le j}
A[k]$. They can be solved in constant time after preprocessing $A$ and storing
a structure using $2n+o(n)$ bits. No accesses to $A$ are needed at query time
\cite{FH11}. The solution requires $lca$ queries on a tree called a 
``2d-min-heap'', and we implement it over our compact trees \cite{SN10}.

{\em Direct Access Codes} \cite{BLN09} represent a sequence of variable-length 
numbers by packing them into chunks of length $b$. Then the chunks are 
rearranged to allow one accessing any $\ell$-bit number in the sequence in 
time $O(\ell/b)$. The space overhead for a number of $\ell$ bits is 
$\ell/b + b$. We use their implementation, which chooses optimally the $b$
values.

{\em Wavelet trees} \cite{GGV03} can be used to represent an $n\times r$ grid 
that contains $n$ points, one per column \cite{MN06}.
The root represents the sequence of coordinates $y_i$ of the points 
in $x$-coordinate order. It only stores a bitmap $B[1,n]$ telling at $B[i]$
whether $y_i < r/2$ or not. Then the points with $y_i < r/2$ 
are represented, recursively, on the left child of the root, and the others
on the right. Adding $rank$ capabilities to the bitmaps, the wavelet tree 
requires overall $n\log r (1+o(1))$ bits and can track any point towards
its leaf (where the $y_i$ value is revealed) in time $O(\log r)$. 
It can also count, in $O(\log r)$ time, the number of 
points lying inside a rectangle $[x_1,x_2] \times [y_1,y_2]$:
Start at the root with the interval $[x_1,x_2]$ and project those values
towards the left and right child (on the left child the interval is
$[rank_0(B,x_1-1)+1,rank_0(B,x_2)]$, and similarly with $rank_1$ on the right).
This is continued until reaching the $O(\log r)$ wavelet tree nodes that
cover $[y_1,y_2]$. Then the answer is the sum of the lengths of the mapped 
intervals $[x_1^i,x_2^i]$. One can also track those points toward the
leaves and report them, each in time $O(\log r)$.
We use a simple balanced wavelet tree without pointers \cite{CN08}.

{\em Muthukrishnan's algorithm} \cite{M02} for listing the distinct elements
in a given interval $A[i,j]$ of an array $A[1,n]$ uses another array 
$C[1,n]$ where $C[i] = \max \{ j<i,~ A[j]=A[i]\} \cup \{-1\}$, which is
preprocessed for range minimum queries. Each value $C[m]<i$ for
$i \le m \le j$ is a distinct value $A[m]$ in $A[i,j]$. A range minimum
query in $C[i,j]$ gives one such value $m$, and then we continue recursively
on $A[i,m-1]$ and $A[m+1,j]$ until the minimum is $\ge i$. One retrieves any
$k$ unique elements in time $O(k)$.

\section{The Optimal-Time Linear-Space Solution}

Our implementation is based on the framework proposed by Hon, Shah and Vitter
\cite{HSV09} and then followed by Navarro and Nekrich \cite{NN12}: Let
$\mathcal{T}$ be the suffix tree for the concatenation $T$ of a collection of
documents $D_1,\dots,D_d$. This tree contains the nodes corresponding to
all the suffix trees $\mathcal{T}_i$ of the documents $D_i$: For
each node $u \in \mathcal{T}_i$, there is a node $v \in \mathcal{T}$ 
such that $path(v) = path(u)$. We will say that $v=map(u,i)$. Also, let
$parent(u)$ be the parent of a node $u$ and $depth(u)$ be its depth.

They store $\mathcal{T}$ plus additional information on the
trees $\mathcal{T}_i$. If $v=map(u,i)$, then they store $i$ in a list
called {\em F-list} associated to $v$. Further, for each $v=map(u,i)$ they
store a pointer $ptr(v,i) = map(parent(u),i)$, noting where the parent
of $u$ maps in $\mathcal{T}$. We add a dummy root $\rho$ to $\mathcal{T}$ so 
that $ptr(v,i)=\rho$ if $u$ is the root of $\mathcal{T}_i$.

Together with the pointers $ptr(v,i)$ they also store a weight $w(v,i)$,
which is the relevance of $path(u)$ in $D_i$. This relevance
can be any function that depends on the set of starting positions of $path(u)$
in $D_i$. In this paper we focus on a simple one: the number of leaves of $u$
in $\mathcal{T}_i$, that is, the {\em term frequency}.

Let $v$ be the locus of $P$. Hon et al.~\cite{HSV09} prove that, for each 
distinct document $D_i$ where $P$ appears, there is exactly one pointer 
$ptr(v'',i)=v'$ going from a descendant $v''$ of $v$ ($v$ itself included)
to a (strict) ancestor $v'$ of $v$, and $w(v'',i)$ is the
relevance of $P$ in $D_i$. Therefore, they find the $k$ largest $w$ values in
this set.

Navarro and Nekrich \cite{NN12} represent this structure as a grid
of size $O(n) \times O(n)$ with labeled weighted 
points, as follows. They traverse $\mathcal{T}$ in preorder. For each node $v
\in \mathcal{T}$, and for each pointer $ptr(v,i)=v'$, they add a new rightmost 
$x$-coordinate 
with only one point, with $y$-coordinate equal to $depth(v')$, weight
equal to $w(v,i)$, and label equal to $i$. At query time, they find the locus
$v$ of $P$, determine the range $[x_1,x_2]$ of all the $x$-coordinates filled
by $v$ or its descendants, find the $k$ highest-weighted points in
$[x_1,x_2] \times [0,depth(v)-1]$, and report their labels. A linear-space
representation (yet with a large constant) allows them to carry out this task in
time $O(m+k)$.

\section{Our Compressed Representation}

We describe our compressed data structures we use and how we carry out
the search.

\paragraph{Suffix tree.}

We use a CSA \cite{BNesa11} requiring $nH_k(T)+o(nH_k(T))+O(n)$ bits, which 
computes $[sp,ep]$ corresponding to $P$ in time $O(m)$. It also computes
any $SA[i]$ in time $O(\log\log n)$. For this sake we use a sampling every
$\log\log n$ positions. In the samples we store not the exact position in $T$
but just the document where it lies. Hence we need $O(n \log d / \log\log n) = 
o(n\log d)$ bits for the sampling.

In practice, we use an off-the-shelf CSA (SSA from {\em PizzaChili} site,
{\tt http:// pizzachili.dcc.uchile.cl}), and add a sparse bitmap 
$D[1,n]$ marking where documents start in $T$: the document corresponding 
to $SA[i]$ is $rank_1(D,SA[i])$, computed in time $O(\log\log n)$. While
this is worse than having the CSA directly return documents, it retains our CSA
other pattern matching functionalities.

We also add $2n+o(n)$ bits to describe the
topology of the suffix tree, using a tree representation that carries 
out most of the operations in constant time \cite{SN10}. Note this is just the
topology, not a full suffix tree, so we need to search using the CSA. 

We also add $2n+o(n)$ bits for an RMQ structure on top of Muthukrishnan's
array $C$ \cite{M02}, which can list $k$ distinct documents in any
interval $SA[sp,ep]$ in time $O(k)$.

\vspace*{-4mm}
\paragraph{Mapping to the grid.}

The grid is of width $\sum_i |\mathcal{T}_i| \le 2n$, as we add one coordinate 
per node in the suffix tree of each document. To save space, we will consider
a {\em virtual} grid just as defined, but will store a narrower {\em physical}
grid. In the physical grid, the entries corresponding to leaves of 
$\mathcal{T}$ (which contain exactly one pointer $ptr(v,i)$) will not be
represented. Thus the physical grid is of width at most $n$. 
This {\em frequency thresholding} is a key idea, as it halves the space of most
structures in our index.

Two bitmaps will be used to map between the suffix array, the suffix tree, 
and the virtual and physical grids: $B[1,2n]$ and $L[1,2n]$. Bitmap $B$ will
mark starting positions of nodes of $\mathcal{T}$ in the physical grid:
each time we arrive at an internal node $v$ we add a 1 to $B$, and each time we 
add a new $x$-coordinate to the grid (due to a pointer $ptr(v,i)$) we add a 0
to $B$. Bitmap $L$ will mark leaves in the preorder traversal of
$\mathcal{T}$, using a 1 for leaves and a 0 for internal nodes.

\vspace*{-4mm}
\paragraph{Representing the grid.}

In the grid there is exactly one point per $x$-coordinate. We represent with
a wavelet tree \cite{GGV03} the sequence of corresponding $y$-coordinates.
Note that the height of this grid is $c.\log n$ for some constant $c$ 
a.s. \cite[Thm.\ 1(ii) and Remark 2(iv)]{Szp93}. 
Thus, the height of the wavelet tree is $\log\log n + O(1)$ and
the wavelet tree requires $n\log\log n (1+o(1))$ bits in total, a.s.\ 
(from now on we will omit, except in the theorems, that our results hold 
almost surely and not in the worst case). 

Each node $v$ of the wavelet tree represents a subsequence of the
original sequence of $y$-coordinates. We consider the (virtual) sequence of 
the weights associated to the points represented by $v$, $W(v)$, and build an 
RMQ data structure \cite{FH11} for $W(v)$. This structure requires 
$2|W(v)|+O(|W(v)|/\log n)$. This adds up to $2n\log\log n (1+o(1))$ for the
whole wavelet tree.

\vspace*{-4mm}
\paragraph{Representing labels and weights.}

The labels of the points, that is, the document identifiers, are represented
directly as a sequence of at most $n\lceil \log d\rceil=n\log d + O(n)$ bits, 
aligned to the bottom of the wavelet tree. Given any point to report, we 
descend to the leaf in $O(\log\log n)$ time and retrieve the document 
identifier.

The weights are stored similarly, but using direct access codes \cite{BLN09}
to take advantage of the fact that most weights (term frequencies) are
small. Note that the subtree size of each $\mathcal{T}_i$ internal node will 
be stored exactly once as the weight of some $ptr(v,i)$.

We analyze now that the number of bits required to store those numbers. 
Let $n_i = |D_i|$. Since the height of any $\mathcal{T}_i$ is $O(\log n_i)$, 
so is the depth of any node. The sum of the depths of all the nodes is then
$O(n_i\log n_i)$, and this is also the sum of all the subtree sizes. 
Distributing those sizes over the $n_i$ nodes uniformly (which gives a pretty 
pessimistic worst case for the sum of the logarithms) gives $O(\log n_i)$ for 
each. Thus the number of bits required to represent the sizes is at most
$\log\log n_i + O(1) \le \log\log n + O(1)$. Using direct access codes with 
block size $b=\sqrt{\log\log n}$ poses an extra overhead of
$O(\sqrt{\log\log n}) = o(\log\log n)$ bits. Hence all the weights can be
stored in $n\log\log n (1+o(1))$ bits and accessed in time 
$O(\sqrt{\log\log n})$.\footnote{We conjecture that the number of bits is
actually $O(n)$, which we can prove only for uniformly distributed texts.}

\vspace*{-4mm}
\paragraph{Answering queries.}

The first step to answer a query is to use the CSA to determine the range 
$[sp,ep]$ in time $O(m)$. To find the locus $v$ of $P$ in the topology of
the suffix tree, we compute $l$ and $r$, the $sp$th and $ep$th 
leaves of the tree, respectively, using $l=preorderselect(select_1(L,sp))$ 
and $r=preorderselect(select_1(L,ep))$, and then we have $v=lca(l,r)$. 
All those operations take $O(1)$ time.

To determine the horizontal extent $[x_1,x_2]$ of the grid that 
corresponds to the locus node $v$, we first compute $p_1=preorder(v)$ and
$p_2=p_1+subtreesize(v)$. This gives the preorder range $[p_1,p_2)$ 
including leaves.
Now $l_1 = rank_1(L,p_1)$ and $l_2 = rank_1(L,p_2-1)$ gives the number of leaves
up to those preorders. Then, since we have omitted the leaves in the
physical grid, we have $x_1 = select_1(B,p_1-l_1)-(p_1-l_1)+1$ and
$x_2 = select_1(B,p_2-l_2)-(p_2-l_2)$. 
The limits in the $y$ axis are just $[0,depth(v)-1]$. Thus the grid area to
query is determined in constant time.

Once the range $[x_1,x_2] \times [y_1,y_2]$ to query is determined, we 
proceed to the grid. We determine the wavelet tree nodes that cover the
interval $[y_1,y_2]$, and map the interval $[x_1,x_2]$ to all of them. 
As there are at most two such nodes per level, there are $O(\log\log n)$ 
nodes covering the interval, and they are found in $O(\log\log n)$ time.

We now use a top-$k$ algorithm for wavelet trees \cite{NR11}.
Let $v_1, v_2, \ldots, v_s$ the wavelet tree nodes that cover $[y_1,y_2]$
and let $[x_1^i,x_2^i]$ be the interval $[x_1,x_2]$ mapped to $v_i$. For each 
of them we compute $\textsc{rmq}_{W(v_i)}(x_1^i,x_2^i)$ to find the position 
$x_i$ with the largest weight among the points in $v_i$, and find out that
weight and the corresponding document, $w_i$ and $d_i$. We set up a max-priority
queue that will hold at most $k$ elements (elements smaller than the $k$th are
discarded by the queue). We initially insert the $O(\log\log n)$ tuples
$(v_i,x_1^i,x_2^i,x_i,w_i,d_i)$, being $w_i$ the sort key. Now we iteratively
extract the tuple with the largest weight, say
$(v_j,x_1^j,x_2^j,x_j,w_j,d_j)$. We report the document $d_j$ with weight $w_j$,
and create two new ranges in $v_j$: $[x_1^j,x_j-1]$ and $[x_j+1,x_2^j]$.
We compute their RMQ, find the corresponding documents and weights, and 
reinsert them in the queue. After $k$ steps, we have reported the
top-$k$ documents.

Using a y-fast trie \cite{Wil83} for the priority queue, the total time is 
$O(\log\log n)$ to find the cover nodes,
$O((\log\log n)^2)$ to determine their tuples and insert them in the queue,
and $O(k\log\log n)$ to extract the minima, 
compute and reinsert new tuples.

We remind that we have not stored the leaves in the grid. Therefore, if the
procedure above yields less than $k$ results, we must complete it
with documents where the pattern appears only once. We use Muthukrishnan's
algorithm \cite{M02} with the RMQ structure on the $C$ array. We extract
distinct documents until we obtain $k$ distinct documents in total, counting
those already reported with the grid. This requires at most $2k$ steps, as we 
can revisit the documents reported with the grid. Each
step requires $O(\log\log n)$ time to compute the document identifier.

\vspace*{-2mm}
\begin{theorem}
Given $d$ documents concatenated into a text $T[1,n]$, 
we can build an index requiring almost surely
$(H_k(T) + \log d + 4\log\log n)(n+o(n))$ bits,
which can report the top-$k$ documents most relevant to a search pattern
$P[1,m]$ in time
$O(m + (k+\log\log n)\log\log n)$ almost surely.
Our structure can be built in time $O(n\log\sigma+n\log\log n)$ 
(details omitted).
\end{theorem}

\no{
\begin{algorithm}[h]
\caption{Top-$k$(P, m, k)}

\begin{algorithmic} 
\STATE $sp,ep \leftarrow locate_{CSA}(P,m)$
\STATE $u  \leftarrow LCA(leaf_{CST}(sp),leaf_{CST}(ep))$
\STATE $p  \leftarrow preorder\_rank(u)$
\STATE $p'  \leftarrow p+subtree\_size(u)$
\STATE $x_0  \leftarrow select_1(p,B) - p+ 1$
\STATE $x_1  \leftarrow select_1(p',B) - p'$
\STATE $d  \leftarrow depth_{CST}(u)-1$
\STATE $documents \leftarrow range\_query_{WT}(x_0,x_1,0,d-1,k)$
\STATE $\mathbf{return}\; documents$ 
\end{algorithmic}
\end{algorithm}
}

\section{Experiments and Results}

We compared our solution to the implementation of Navarro and Valenzuela \cite{NV12}, which is the current state of the art. We use various compact data structures implementations from {\em libcds} ({\tt http://libcds.recoded.cl}). 
We used the following collections in our experiments. Their grid heights are
between 5 and 9.




\vspace{-2mm}
\begin{itemize}
\item[] \textbf{\emph{DNA.}} A sequence of 10,000 highly repetitive ($0.05\%$ difference between documents) synthetic DNA sequences with 100,030,004 bases in total. \vspace{-10pt}
\item[] \textbf{\emph{KGS.}} A collection of 18,383 sgf-formatted Go game records from year 2009 ({\tt http://www.u-go.net/gamerecords}), containing 26,351,161 chars. \vspace{-10pt} 
\item[] \textbf{\emph{Proteins.}} A collection of 143,244 sequences of Human and Mouse Proteins ({\tt http://www.ebi.ac.uk/swissprot}), containing 59,103,058 symbols. 
\vspace{-10pt} 
\item[] \textbf{\emph{FT91-94.}} A sample of 40,000 documents from TREC Corpus FT91 to 94 ({\tt http://trec.nist.gov}) containing 93,498,090 characters.  \vspace{-10pt}
\item[] \textbf{\emph{Wikipedia.}} A sample of 40,000 documents from the English Wikipedia containing 83,647,329 characters.
\end{itemize}

\vspace*{-2mm}
The experiments were performed in an Intel(r) Xeon(r) model E5620 running 
at 2.40 GHz with 96GB of RAM and 12,288KB cache. The operating system is Linux with kernel 2.6.31-41 64 bits and we used the GNU C compiler version 4.4.3 with -O3 optimization parameter. For queries, we selected 4,000 random substrings of length 3 and 8, and obtained the top-$k$ documents for each, for $k = 10..100$ every 10 values.

\begin{figure}[t]
\centering
\begin{tabular}{ll}
 \includegraphics[scale=0.28,angle=270]{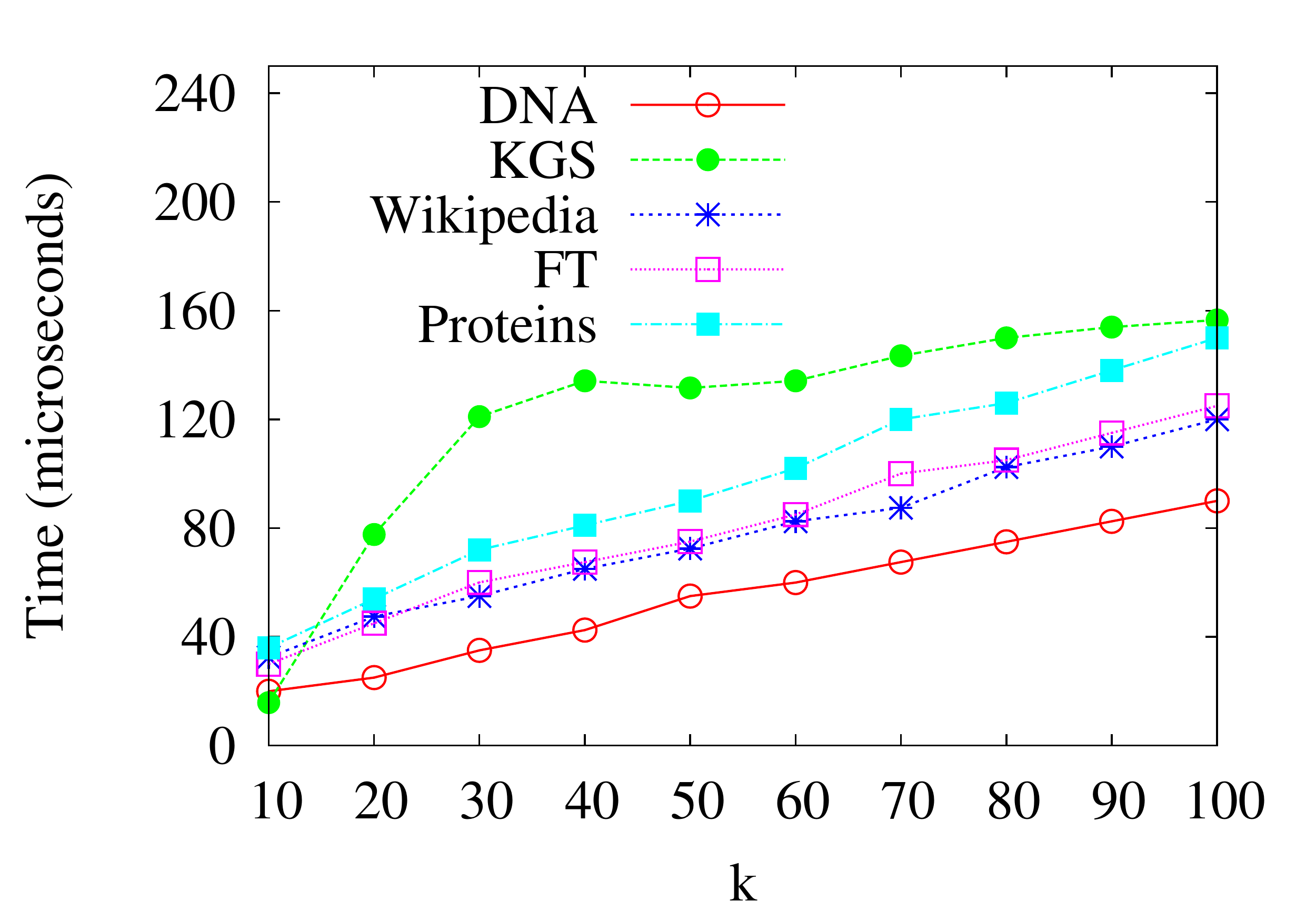} & \includegraphics[scale=0.28,angle=270]{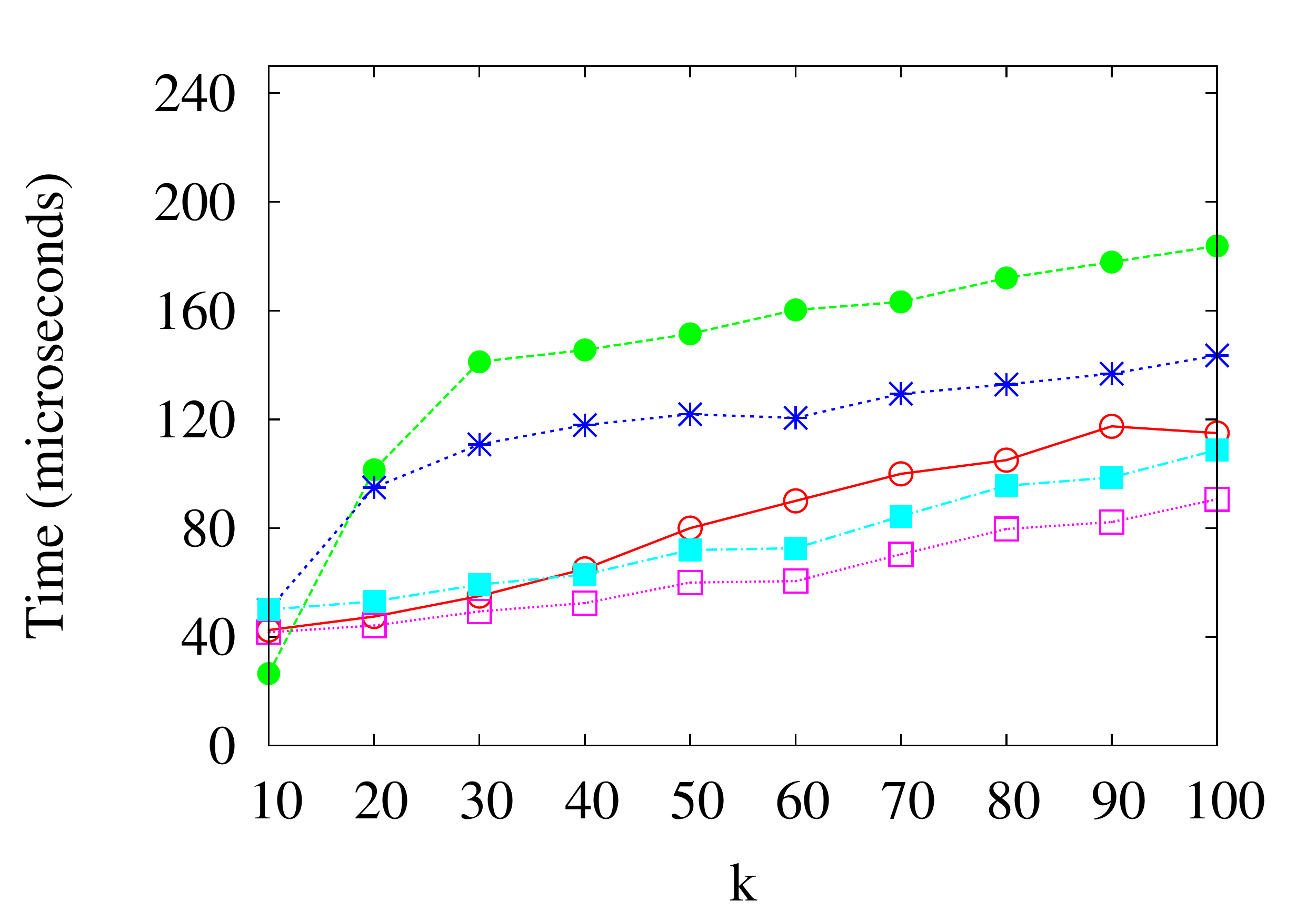}
\end{tabular}

\vspace*{-3mm}
\caption{Time performance as a function of $k$, for $m=3$ (left) and $m=8$ (right).}
\label{fig:time}
\end{figure}


\begin{figure}[t]

\centering
\begin{tabular}{ll}

 \includegraphics[scale=0.28,angle=270]{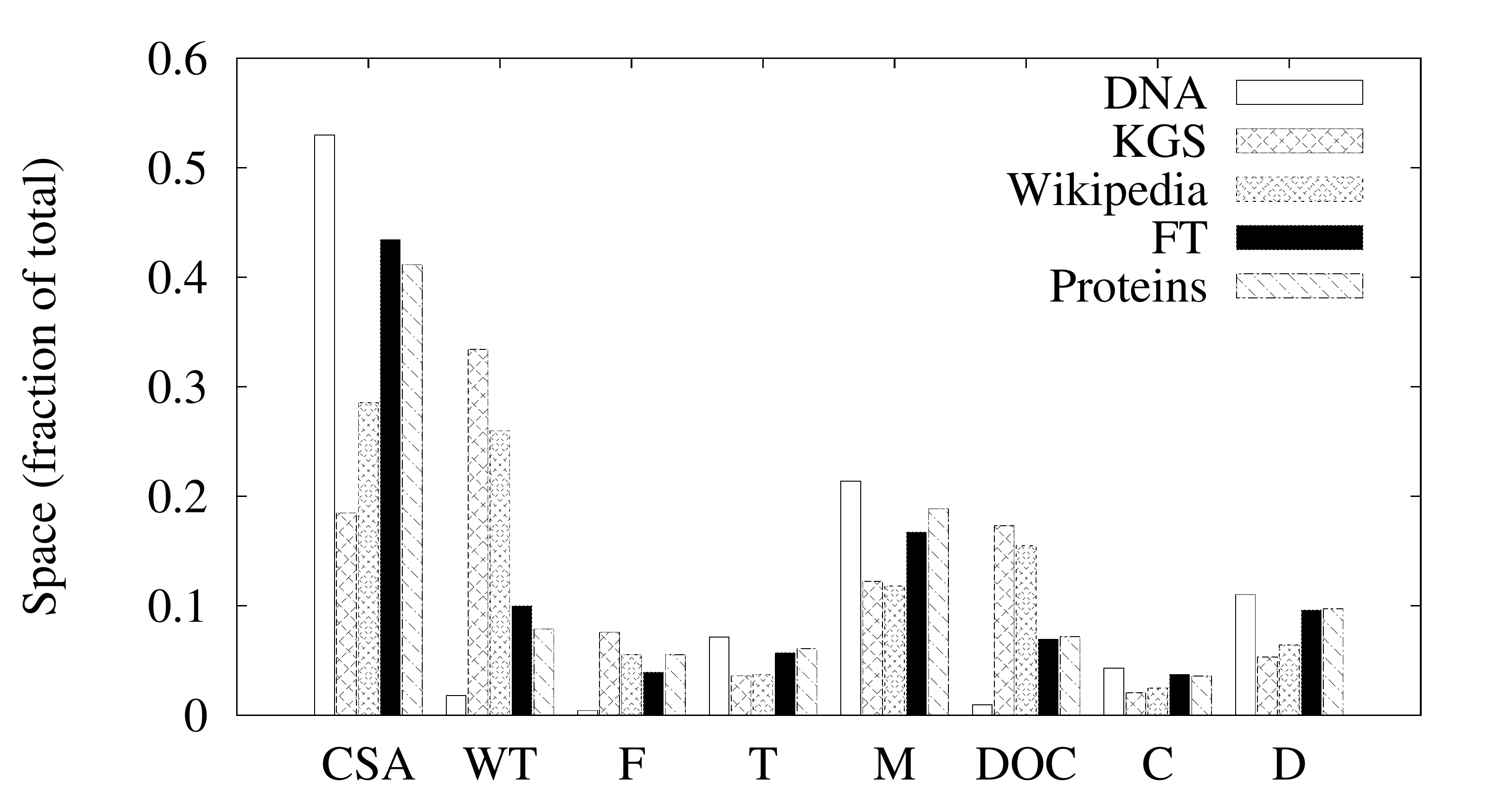} & \includegraphics[scale=0.28,angle=270]{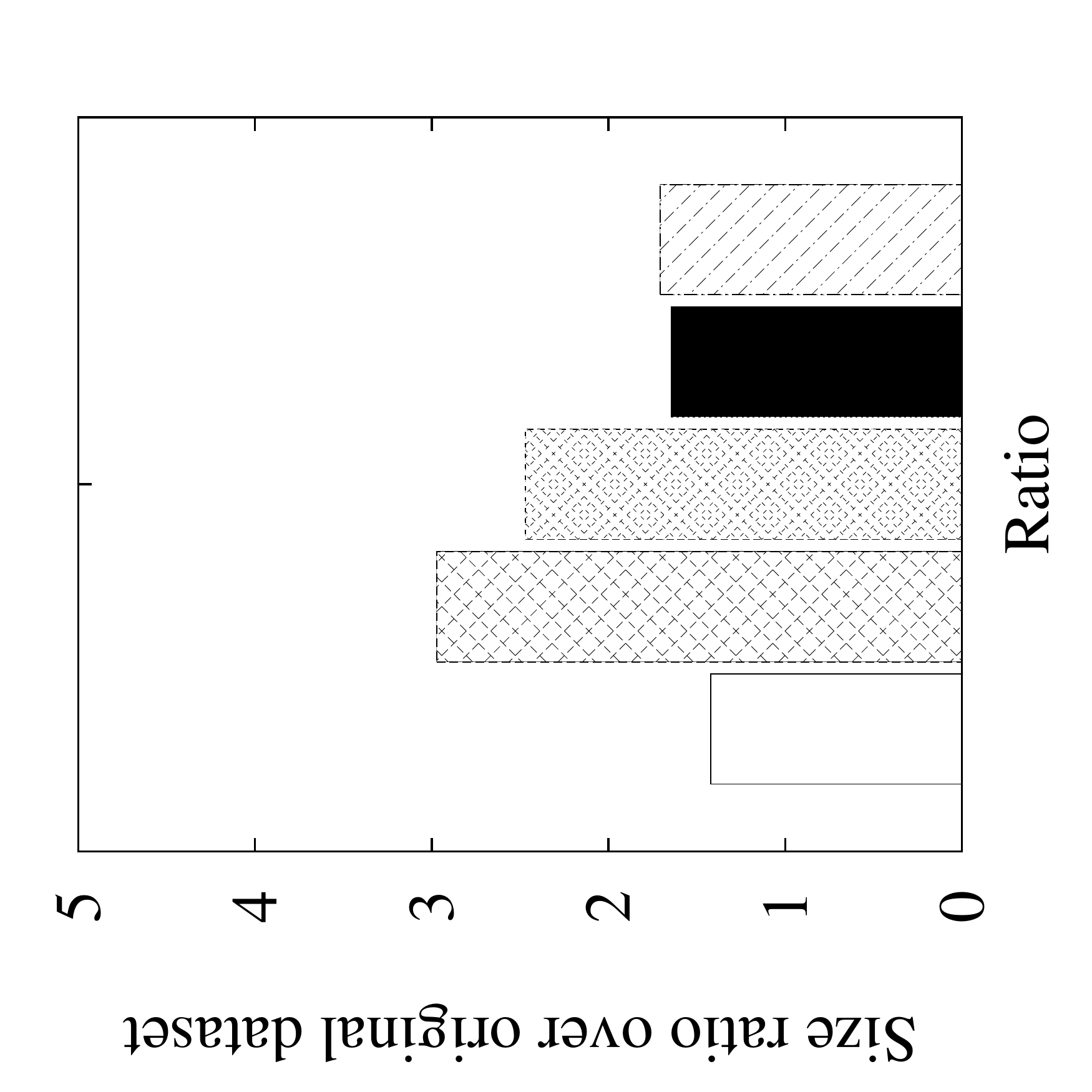}
\end{tabular}

\vspace*{-2mm}
\caption{Space consumption of the different compact data structures employed (left) and the size ratio over the original dataset for the different collections (right).}
\label{fig:space}
\end{figure}

Figure~\ref{fig:time} shows time performance as a function of $k$. The time taken by the CSA search is always near 20 microseconds, after which the index takes about $k$ microseconds. In some cases ({\em KGS}, or {\em Wikipedia} for $m=8$)
there are no enough results with frequency larger than 1, and document listing
must be activated, which slows down the process to $1.6k$--$4k$ microseconds.
Note that in practice one may wish to avoid listing those low-frequency
documents anyway.

Figure~\ref{fig:space} (left) shows the fraction of space used by the different data structures employed: the CSA, the augmented wavelet tree (WT), the DAC-encoded frequencies (F), the suffix tree topology (T), the document identifiers (DOC), the mapping bitmaps $B$ and $L$ (M), the RMQ structure for Muthukrishnan's document listing (C), and the sparse bitmap $D$ marking document limits.
Figure~\ref{fig:space} (right), shows the size ratio over the original dataset (considering one byte per symbol). 
The values vary between 1.5 and 3 times the size of the collection. Note that,
within this space, {\em we can reproduce any document of the collection}, as 
our CSA offers access to them.

\vspace*{-2mm}
\section{Final Remarks}

\begin{table}
\footnotesize
\begin{center}
\begin{tabular}{|l||r|r|r|r||r|r|r|r|}
\hline
           & \multicolumn{4}{c||}{vs compressed \cite{NV12}} 
           & \multicolumn{4}{c|}{vs uncompressed \cite{NV12}}  \\
\cline{2-9}
Collection & $\times$ space & \multicolumn{3}{c||}{speedup} 
           & $\times$ space & \multicolumn{3}{c|}{speedup} \\
\cline{3-5} \cline{7-9}
           &                & $m$ & $k=10$ & $k=100$  
           &                & $m$ & $k=10$ & $k=100$ \\
\hline
KGS & 0.90 
 & 3 & 13.6 & 13.0 
    & 0.85
 & 3 & 3.9 & 5.8 \\
&& 8 & 24.3 & 26.1 
&& 8 & 3.1 & 5.3 \\ 
\hline
Wikipedia & 1.05 
 & 3 & 4.2 & 4.4 
    & 0.79
 & 3 & 2.5 & 1.0 \\
&& 8 & 6.4 & 5.9 
&& 8 & 3.4 & 3.6 \\
\hline
Proteins & 0.53 
 & 3 &  2.8  & 1.1  
    & 0.41 
 & 3 & 22.0 & 12.3 \\
&& 8 & 28.3 & 2.2 
&& 8 & 2.0 & 0.8 \\
\hline
\end{tabular}
\vspace{-10pt}
\end{center}
\caption{Comparison to the best previous work \cite{NV12}, giving the fraction of their space we use, and the speedup we obtain with respect to them.}
\label{tab:comp}
\end{table}

Table~\ref{tab:comp} compares our solution with previous work \cite{NV12} on 
the three collections shared, taking their best compressed (from their variant
{\em WT-Alpha+SSGST} plus more recent improvements) and uncompressed (from their variant {\em WT-Plain+SSGST}) results.\footnote{In their paper \cite{NV12},
the CSA space is not included. We have added ours for a fair comparison.}
Our structure is at most only 5\% larger. When both use about the same space,
our structure is 4 to 25 times faster. In other cases our structure can use
up to half the space, and it is still faster, up to 3 times (for large $k$ and
$m$ we must resort to much document listing, where their wavelet tree on
documents is faster).

Needless to say, this is a remarkable result for a structure that, in theory
\cite{NN12}, used about 80 times the collection size. We have sharply compressed
it while retaining the best ideas that led to its optimal time. We believe this
establishes a new direction in which research on space-efficient top-$k$ retrieval could be focused: Rather than sampling the suffix tree nodes \cite{HSV09,NV12},
threshold the document {\em frequencies} we store (curiously, this is closer in spirit to the first, superlinear-size, proposed top-$k$ solution
\cite{HSW09,HPSW10}). For example, can we discard all the frequencies 
below a threshold $f$ and efficiently list them if needed? Our work shows
this is possible at least for $f=1$. 

Our approach easily extends to relevance functions other than term frequency.
In most cases it is sufficient to store the appropriate weights in our data
structure. Even if these are not compressible, the space should not grow
up too much. Our structure also trivially solves other document listing 
problems, like $k$-mining (list the documents where $P$ appears at least $k$ 
times). Muthukrishnan \cite{M02} solves it in optimal time $O(m+occ)$ and $O(n\log n)$
bits for $k$ fixed at indexing time. For variable $k$ the space is $O(n\log^2 n)$. Our compressed structure, without modifications, solves both variants in time $O(m+(occ+\log\log n)\log\log n)$.


\end{document}